# Alcohols on the Rocks: Solid-State Formation in a H$_3$CC≡CH + OH Cocktail under Dark Cloud Conditions

Danna Qasim,*[,†] Gleb Fedoseev,[‡] Thanja Lamberts,[§] Ko-Ju Chuang,[†,#] Jiao He,[†] Sergio Ioppolo,[∥] Johannes Kästner,[⊥] and Harold Linnartz[†]

[†]Sackler Laboratory for Astrophysics, Leiden Observatory, Leiden University, P. O. Box 9513, NL-2300 RA Leiden, The Netherlands

[‡]INAF−Osservatorio Astrofisico di Catania, via Santa Sofia 78, 95123 Catania, Italy

[§]Leiden Institute of Chemistry, Leiden University, P. O. Box 9502, NL-2300 RA Leiden, The Netherlands

[∥]School of Electronic Engineering and Computer Science, Queen Mary University of London, Mile End Road, London E1 4NS, United Kingdom

[⊥]Institute for Theoretical Chemistry, University of Stuttgart, 70569 Stuttgart, Germany

Ⓢ Supporting Information

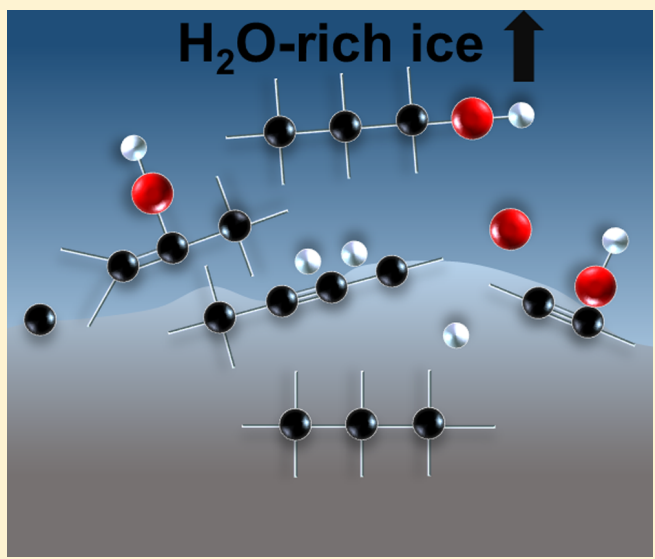

**ABSTRACT:** A number of recent experimental studies have shown that solid-state complex organic molecules (COMs) can form under conditions that are relevant to the CO freeze-out stage in dense clouds. In this work, we show that alcohols can be formed well before the CO freeze-out stage (i.e., during the very early stage of the H$_2$O-rich ice phase). This joint experimental and computational investigation shows that isomers *n*-propanol and isopropanol (H$_3$CCH$_2$CH$_2$OH and H$_3$CCHOHCH$_3$) and *n*-propenol and isopropenol (H$_3$CCH=CHOH and H$_3$CCOH=CH$_2$) can be formed in radical-addition reactions starting from propyne (H$_3$CC≡CH) + OH at the low temperature of 10 K, where H$_3$CC≡CH is one of the simplest representatives of stable carbon chains already identified in the interstellar medium (ISM). The resulting average abundance ratio of 1:1 for *n*-propanol:isopropanol is aligned with the conclusions from the computational work that the geometric orientation of strongly interacting species is influential to the extent of which "mechanism" is participating and that an assortment of geometries leads to an averaged-out effect. Three isomers of propanediol are also tentatively identified in the experiments. It is also shown that propene and propane (H$_3$CCH=CH$_2$ and H$_3$CCH$_2$CH$_3$) are formed from the hydrogenation of H$_3$CC≡CH. This experimental finding falls in line with the lower activation barrier of hydrogenation of a C=C bond in comparison to a C≡C bond. Reactants and products are probed by temperature-programmed desorption−quadrupole mass spectrometry (TPD-QMS) and reflection absorption infrared spectroscopy (RAIRS). Product relative abundances are determined from TPD-QMS data. Computationally derived activation barriers give additional insight into what types of reactions and mechanisms are more likely to occur in the laboratory and in the ISM. Our findings not only suggest that the alcohols studied here share common chemical pathways and therefore can show up simultaneously in astronomical surveys but also that their extended counterparts that derive from polyynes containing H$_3$C−(C≡C)$_n$−H structures may exist in the ISM. Such larger species, such as fatty alcohols, are the possible constituents of simple lipids that primitive cell membranes on the early Earth are thought to be partially composed of.

*continued...*

KEYWORDS: *astrochemistry, astrobiology, solid-state, interstellar, ice, complex organic molecules, infrared, mass spectrometry*

## 1. INTRODUCTION

The origin of cosmic carbon lies in the outflows of carbon-rich stars.[1,2] How carbon evolves into hydrocarbon species, from small molecules such as methane ($CH_4$) and acetylene ($C_2H_2$) to polycyclic aromatic hydrocarbons (PAHs), to carbon nanoparticles, or other carbon-containing species such as alcohols, is far from understood. In addition to "bottom-up" approaches, which merge smaller precursors into larger species (in the gas-phase or solid-state), also "top-down" approaches have been proposed. Both scenarios are considered as likely road maps toward molecular complexity in space, but as mentioned, many details are lacking.[3,4]

From a bottom-up perspective, simple species such as $C_2H_2$ can polymerize to form polyacetylene ($[C_2H_2]_n$).[5,6] The complexity can increase by addition of a methyl (-$CH_3$) group to form methylpolyacetylene. Such species can eventually accrete onto carbonaceous dust grains that were formed from nucleation of PAHs.[2,7] An alternative is that these species form through surface reactions in the ice layers that are on top of dust grains. In this article, we focus on the simplest representative of the latter row of species: methylacetylene, also known as propyne ($H_3CC\equiv CH$). $H_3CC\equiv CH$ has been detected in space.[8-16] Not only is it reported to be observed toward carbon-rich stars,[13] as expected, but it is also reported to be detected toward cold, dark, and dense clouds,[9,12] where it has a column density of $(4-8) \times 10^{13}$ cm$^{-2}$ in TMC-1.[9] Assuming an $H_2$ abundance of $\sim 10^{21}$ cm$^{-2}$,[3] this leads to <1% with respect to $H_2O$ ice following the calculation from Herbst and van Dishoeck.[17]

$H_2O$ ice is formed on dust grains.[18-23] Laboratory experiments show that every formed $H_2O$ molecule, in the solid-state and gas-phase, has an OH radical and/or ion as an intermediate.[24] In the ice, OH radicals have the chance to react with other species that are in direct proximity prior to hydrogenation to yield $H_2O$, as the accretion of an H atom onto a dust grain of radius $10^{-5}$ cm happens once a day.[25] An example of this is the formation of $CO_2$ from CO and OH in $H_2O$-rich ices.[26] Thus, the interaction between OH radicals and $H_3CC\equiv CH$ is a valid topic to be addressed in astrochemical laboratories. This interaction may result in the formation of simple alcohols and polyalcohols.

The formation of alcohols is particularly intriguing as they may have a role in astrobiology, assuming that alcohols can be delivered to planetary bodies such as the early Earth.[27] Simple alcohols are amphiphilic molecules, i.e., both sides of the molecule have different affinities from each other.[28] Because they are composed of a polar head (-OH group) and a hydrophobic tale (aliphatic group), they can take part in the formation of micelles in primordial oceans. More complex alcohols (e.g., polyalcohols and fatty alcohols), in turn, can play a role in the formation of primitive lipids. Like sugars in saccharolipids, sphingosines in sphingolipids, and glycerol in modern phospholipids, fatty alcohols can act as the backbone to which fatty acids are attached. Their presence during abiogenesis is supported by the idea that complex lipids may not have been available on the early Earth,[29,30] and also through the finding that primary alcohols are components of archaea cell membranes.[31]

The mechanism proposed in this work results in the formation of alcohols already in $H_2O$-rich ices, meaning, the alcohols can be synthesized before the CO freeze-out stage, well below extinctions ($A_V$) of 9 in dense atomic/molecular cores.[32] In the literature, most solid-state laboratory experiments report the formation of simple alcohols (e.g., methanol ($CH_3OH$) and ethanol ($H_3CCH_2OH$)) and polyalcohols (e.g., ethylene glycol ($HOCH_2CH_2OH$), glycerol ($HOCH_2CHOHCH_2OH$), and methoxymethanol ($H_3COCH_2OH$)) in the context where much CO has already been frozen out.[33-40] These experiments can be divided into two subgroups: "energetic" and "non-energetic" processing, where "non-energetic" refers to a radical-induced process without the involvement of UV, cosmic rays, and/or other "energetic" particles.[41] In the "energetic"-induced studies, $CH_3OH$- and/or CO-containing ices are irradiated to form alcohols, and $CH_3OH$ is either mixed with CO or explicitly stated to be a product of CO hydrogenation.[33-36] In experiments that focus on "non-energetic" processes, alcohols are formed by reactions that involve the hydrogenation of a CO-rich ice.[37-40]

Many of the icy alcohols that have been formed in the laboratory have also been detected as gas-phase species in the interstellar medium (ISM). This includes the detection of $CH_3OH$, $H_3CCH_2OH$, vinyl alcohol ($H_2CCHOH$), $HOCH_2CH_2OH$, and $H_3COCH_2OH$, which were first reported by Ball et al.,[42] Zuckerman et al.,[43] Turner and Apponi,[44] Hollis et al.,[45] and McGuire et al.,[46] respectively. $CH_3OH$ also has been detected in the solid-state.[47] Much effort has been recently put into explaining the transition from frozen to gas-phase $CH_3OH$ to explain, for example, $CH_3OH$ abundances observed in protoplanetary disks.[48] Other alcohols are still elusive, such as $n$-propanol ($H_3CCH_2CH_2OH$) (Qasim et al., submitted for publication).

This work overviews the reaction of solid-state $H_3CC\equiv CH$ with H and OH under conditions relevant to the dense cloud stage. Section 2 provides details on the experimental and computational parameters used for this study. Section 3 presents the findings from the laboratory experiments. Section 4 reports the computationally derived energies that are applicable to the reactions taking place in the experiments. These results are combined with the laboratory work to uncover the products formed and their formation pathways. Section 5 discusses how the formation of such icy alcohols can take place at the interface of astrochemical and astrobiological environments. Finally, the conclusions of this study are summarized in section 6.

Given the many different species that will be discussed in the next sections, Table 1 is added and summarizes the names and chemical structure formulas of all relevant species. A majority of the listed chemicals are the expected products of solid-state $H_3CC\equiv CH$ hydrogenation or hydroxylation.

## 2. METHODOLOGY

**2.1. Experimental Apparatus.** The creation of ices and the subsequent measurements occur within an ultrahigh vacuum (UHV) apparatus, SURFRESIDE$^2$. The main chamber reaches a base pressure of low $10^{-10}$ mbar. Near the center of the chamber, ices (typically tens of monolayers thick) are formed on a gold-plated copper substrate that is attached to a closed cycle helium cryostat. The inclusion of resistive heating and a sapphire rod allows the sample to have a temperature range of 7–450 K. The temperature is measured by a silicon diode sensor that has an absolute accuracy of 0.5 K. Further details of the initial design of SURFRESIDE[2] are found in Ioppolo et al.,[49] and recent upgrades are found in Qasim et al.[50] and Chuang et al.[51]

Table 1. Chemical Terminology Used in This Article

| IUPAC name | referred to in this article |
|---|---|
| propan-1-ol ($H_3CCH_2CH_2OH$) | n-propanol |
| propen-1-ol ($H_3CCH=CHOH$) | n-propenol |
| propan-2-ol ($H_3CCHOHCH_3$) | isopropanol |
| propen-2-ol ($H_3CCOH=CH_2$) | isopropenol |
| propane-1,1-diol ($H_3CCH_2CH(OH)_2$) | propane-1,1-diol |
| propane-2,2-diol ($H_3CC(OH)_2CH_3$) | propane-2,2-diol |
| propane-1,2-diol ($H_3CCHOHCH_2OH$) | propane-1,2-diol |
| propan-2-one ($H_3COCH_3$) | acetone |
| propanal ($H_3CCH_2CHO$) | propanal |
| propanoic acid ($H_3CCH_2COOH$) | propanoic acid |
| propyne ($H_3CC\equiv CH$) | $H_3CC\equiv CH$ |
| propene ($H_3CCH=CH_2$) | $H_3CCH=CH_2$ |
| propane ($H_3CCH_2CH_3$) | $H_3CCH_2CH_3$ |

Two atomic beamlines are connected to the main chamber: a hydrogen atom beam source (HABS) and a microwave atom source (MWAS). In this study, only the HABS is used, and more details about the design of the source is found in Tschersich and Von Bonin,[52] Tschersich,[53] and Tschersich et al.[54] The HABS chamber is also under UHV conditions, where it reaches a base pressure of low $10^{-10}$ mbar. To form hydrogen atoms, hydrogen molecules (Linde 5.0) are thermally cracked by heated tungsten. This process also increases the kinetic energy of the H atoms. To cool these atoms to room temperature, a nose-shaped quartz tube is positioned at the exit of the HABS source, which allows excess energy to be transferred via collisions with glass walls. Upon impact with the icy surface, a fraction of the impinging H atoms temporarily sticks to the ice that covers the surface and is thermalized. These H atoms are then available for the reactions through the Langmuir–Hinshelwood mechanism. This mechanism was confirmed in several studies, where the initial step in the reaction chain initiated by H atoms exhibits a significant activation barrier and requires quantum tunneling to proceed.[39,41,50,55−57]

All gases and vapors are prepared within a turbomolecularly pumped gas manifold. $H_3CC\equiv CH$ (Sigma-Aldrich, 97%), $O_2$ (Linde Gas 99.999%), and $^{18}O_2$ (Campro Scientific, 97%) gases enter the main chamber through one of two dosing lines that are each connected to manually operated leak valves. n-Propanol (Honeywell, 99.9%) and isopropanol (Sigma-Aldrich, 99.8%) are placed in a tube and freeze–pump–thawed in order to rid them of volatile impurities.

Two techniques are used to examine ice constituents and consequently the underlying ice chemistry: reflection absorption infrared spectroscopy (RAIRS) and temperature-programmed desorption–quadrupole mass spectrometry (TPD-QMS). In this study, RAIRS is specifically exploited to identify the species formed at 10 K in situ. Spectra are recorded by a Fourier transform infrared (FTIR) spectrometer that utilizes a wave-number range of 700–4000 cm$^{-1}$ and can ultimately span to 6000 cm$^{-1}$. A resolution of 1 cm$^{-1}$ is chosen. Vibrational mode assignments in the RAIR spectra originate from the NIST database.[58]

TPD-QMS is additionally utilized to probe newly formed ice species—particularly species that present a number of unresolved and/or overlapping infrared peaks. Employment of a QMS with an electron impact ionization energy of 70 eV allows comparison of the fragmentation (dissociative ionization) patterns from the experiments to fragmentation patterns found in the NIST database.[59] The relative abundances of $H_3CC\equiv CH$, $H_3CCH=CH_2$, and $H_3CCH_2CH_3$, as well as n-propanol and isopropanol, are determined by a combination of the TPD-QMS data recorded at a molecule specific temperature and mass spectrometry data from NIST. The formula used to determine their relative abundances can be found in Martín-Doménech et al.[60] $H_3CC\equiv CH$, $H_3CCH=CH_2$, and $H_3CCH_2CH_3$ have similar ionization cross-sections of $7.66 \times 10^{-16}$, $8.74 \times 10^{-16}$, and $8.62 \times 10^{-16}$ cm$^2$, respectively.[61] Additionally, the QMS sensitivity values of their correlating mass fragments, $m/z = 43$, $m/z = 42$, and $m/z = 41$, respectively, are similar.[62] Therefore, only the fragmentation factors and relative intensities are taken into account, where the relative intensities are measured in the temperature range of 70–110 K. This method can also be applied to determining the relative abundance of n-propanol and isopropanol in the temperature range of 120–190 K, using $m/z$ values of 31 for n-propanol and 45 and 59 for isopropanol. However, their respective sensitivity values of 0.3179, 0.1762, and 0.0982 are taken into account, as their values will significantly influence the determined relative abundances. A TPD ramp rate of 5 K/min is applied to all experiments.

Table 2. List of Experiments Performed and the Corresponding Experimental Parameters[a]

| no. | exp | $T_{sample}$, (K) | flux$_{HC\equiv CCH}$, (cm$^{-2}$ s$^{-1}$) | flux$_H$, (cm$^{-2}$ s$^{-1}$) | flux$_{O_2}$, (cm$^{-2}$ s$^{-1}$) | flux$_{other}$, (cm$^{-2}$ s$^{-1}$) | time, (s) |
|---|---|---|---|---|---|---|---|
| | | | $H_3CC\equiv CH$ Hydrogenation | | | | |
| 1.0 | $H_3CC\equiv CH$ | 10 | $2 \times 10^{12}$ | | | | 21600 |
| 1.1 | $H_3CC\equiv CH$ + H | 10 | $2 \times 10^{12}$ | $5 \times 10^{12}$ | | | 21600 |
| 1.2 | $H_3CC\equiv CH$ + H | 10 | $7 \times 10^{12}$ | $5 \times 10^{12}$ | | | 7200 |
| | | | $H_3CC\equiv CH$ and $O_2/^{18}O_2$ Hydrogenation | | | | |
| 2.0 | $H_3CC\equiv CH$ + H + $O_2$ | 10 | $7 \times 10^{12}$ | $5 \times 10^{12}$ | $1 \times 10^{12}$ | | 21600 |
| 2.1 | $H_3CC\equiv CH$ + H + $^{18}O_2$ | 10 | $7 \times 10^{12}$ | $5 \times 10^{12}$ | $1 \times 10^{12}$ | | 21600 |
| 2.2 | $H_3CC\equiv CH$ + H + $O_2$ | 10 | $2 \times 10^{12}$ | $5 \times 10^{12}$ | $4 \times 10^{12}$ | | 21600 |
| 2.3 | $H_3CC\equiv CH$ + H + $^{18}O_2$ | 10 | $2 \times 10^{12}$ | $5 \times 10^{12}$ | $4 \times 10^{12}$ | | 21600 |
| | | | Reference Experiments | | | | |
| 3.0 | n-propanol | 10 | | | | $3 \times 10^{12}$ | 3600 |
| 3.1 | isopropanol | 10 | | | | $3 \times 10^{12}$ | 3600 |
| 3.2 | isopropanol + H + $O_2$ | 10 | | $5 \times 10^{12}$ | $1 \times 10^{12}$ | $2 \times 10^{10}$ | 7200 |

[a]Fluxes are calculated by the Hertz–Knudsen equation, and the H-flux is derived from Ioppolo et al.[49] "Other" refers to either n-propanol or isopropanol.

**2.2. Experimental Procedure.** The experiments and experimental parameters used in this study are listed in Table 2. Fluxes are determined by the Hertz−Knudsen equation[63] except for the H atom flux, which is based on an absolute D atom flux measurement that is reported in Ioppolo et al.[49] Motivation for the listed experiments is discussed below.

Experiments 1.0−1.2 are used to show what products are formed from the hydrogenation of $H_3CC \equiv CH$ ice, as well as to determine the relative abundance of the newly formed products. The addition of oxygen in experiments 2.0−2.3 is used to study the products formed from $H_3CC \equiv CH$ + OH and their subsequent relative abundance. Note that OH radicals are effectively formed from H + $O_2$.[22] $^{18}O_2$ is used in experiments 2.1 and 2.3 in order to confirm the identity of species formed in experiments 2.0 and 2.2, respectively, by observation of the isotopic shift in the TPD-QMS data. To further confirm the identity of the species formed in experiments 2.0−2.3, the TPD-QMS data of experiments 3.0−3.2 are used as references.

**2.3. Computational Details.** We calculate activation energies and reaction energies for the reactions of hydrogen atoms and hydroxyl radicals with both $H_3CC \equiv CH$ and $H_3CCH = CH_2$ molecules. Benchmark calculations and additional supporting information are found in Supporting Information section S2. Since radicals may attack either the center or exterior carbon atom, this results in a total of eight reactions. Additionally, two isomerization reactions are studied, namely, the conversion from n-propenol to propanal (and vice versa) and isopropenol to acetone (and vice versa).

The potential energy surface (PES) or electronic structure is described by density functional theory (DFT). Following the benchmark calculations performed by Kobayashi et al.,[64] the MPWB1K functional[65] in combination with the basis set def2-TZVP[66] is chosen. The energy and gradient calculations are carried out in NWChem version 6.6.[67] An additional benchmark is performed for the activation energies with the M06-2X functional[68] with the same basis set (def2-TZVP). Furthermore, the interaction energies of the OH−$C_3H_n$ pre-reactive complexes calculated with MPWB1K/def2-TZVP are compared to single-point energies calculated with CCSD(T)-F12/cc-VDZ-F12[69−74] in Molpro version 2012.[75]

Geometry optimizations are carried out for the separated reactant, product, and transition structures and verified by the appropriate number of imaginary frequencies. A transition structure is characterized by the Hessian bearing exactly one negative eigenvalue. To confirm that the found transition structure connects the desired reactant and product, an intrinsic reaction coordinate (IRC) search is conducted. From the end point of the IRC, a reoptimization is performed to obtain the pre-reactive complex (PRC). All calculations are performed with DL-find[76] within Chemshell.[77,78] IRC searches are performed using the algorithm described by Meisner et al.[79] and Hratchian and Schlegel.[80] Finally, for the reaction OH + $H_3CCH = CH_2$, transition states are only found through a nudged elastic band (NEB) approach.

All calculated energies include a zero-point energy (ZPE) correction that is listed separately. Note that these ZPE corrections can be quite sizable and are thus important to include due to their impact on the total activation energy. Finally, activation energies are calculated with respect to both separated reactants (SR) and the pre-reactive complex (PRC). Although the difference between these two approaches lies only in the considered starting point of the reaction, the effect on the activation energies again can be quite pronounced.

All calculations are performed in the gas-phase since we expect the influence of $H_2O$ molecules in the neighborhood of unsaturated hydrocarbons to play a minor role in altering the reaction potential energy landscape.[64]

## 3. EXPERIMENTAL RESULTS AND DISCUSSION

**3.1. Hydrogenation of $HC \equiv CCH_3$ To Form $H_3CCH = CH_2$ and $H_3CCH_2CH_3$.** The formation of $H_3CCH = CH_2$ and $H_3CCH_2CH_3$ by the hydrogenation of $H_3CC \equiv CH$ at 10 K is visible from the RAIR data displayed in Figure 1 (upper panel), and the corresponding vibrational mode assignments are listed in Table 3. Despite the overlap of a number of vibrational bands between $H_3CCH = CH_2$ and $H_3CCH_2CH_3$, there are distinct peaks that are characteristic to these species and also do not overlap with the IR signatures of $H_3CC \equiv CH$. The C=C stretch of $H_3CCH = CH_2$ at 1644 $cm^{-1}$[81] is conveniently isolated, and the C−C stretch of $H_3CCH_2CH_3$ at 851 $cm^{-1}$ is clearly visible. The hydrogenation of $H_3CC \equiv CH$ to form $H_3CCH = CH_2$ and $H_3CCH_2CH_3$ parallels the hydrogenation of the two-carbon counterpart, acetylene (HCCH), which results in the formation of ethene ($H_2C = CH_2$) and ethane ($H_3CCH_3$).[64]

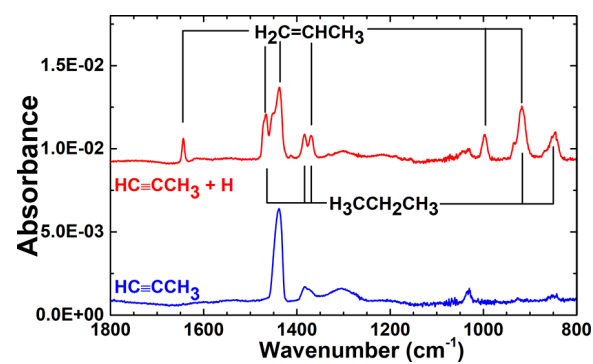

**Figure 1.** RAIR spectra acquired after deposition of $HC \equiv CCH_3$ (bottom spectrum; exp 1.0) and $HC \equiv CCH_3$ + H (top spectrum; exp 1.1) on a 10 K surface. The infrared peaks of newly formed $H_3CCH = CH_2$ and $H_3CCH_2CH_3$ are highlighted. RAIR spectra are offset for clarity.

TPD-QMS spectra provide additional proof for the newly formed $H_3CCH = CH_2$ and $H_3CCH_2CH_3$ from $H_3CC \equiv CH$ hydrogenation and are presented in Figure 2. The m/z fragment values upon 70 eV electron impact ionization with the highest relative intensities for $H_3CCH = CH_2$ and $H_3CCH_2CH_3$ are 41 and 29, respectively (NIST). The desorption peak temperature of $H_3CCH_2CH_3$ from an amorphous solid $H_2O$ surface is ∼80 K,[85] a peak that is also observed in Figure 2. Since $H_3CCH_2CH_3$ is composed of single bonds and $H_3CCH = CH_2$ has a double bond, $H_3CCH_2CH_3$ should have a lower desorption energy than $H_3CCH = CH_2$ (an effect of π stacking).[86] Thus, the desorption at 81 K is assigned as the main desorption peak of $H_3CCH_2CH_3$, and the higher temperature signal at 86 K must be the main desorption peak of $H_3CCH = CH_2$. Additionally, it is observed that some $H_3CCH_2CH_3$ co-desorbs with $H_3CCH = CH_2$ at 86 K, which may be due to the amorphous to crystalline phase transition of $H_3CC \equiv CH$, since m/z = 29 is not a fragment value of H$_3$CCH=CH$_2$ (or H$_3$CC≡CH), yet there is a desorption peak for $m/z$ = 29 at 86 K. The desorption of the bulk of unreacted H$_3$CC≡CH ice peaks around 104 K.

Table 3. Relevant Normal Vibrational Modes Detected in the HC≡CCH$_3$ + H (Experiment 1.1) and HC≡CCH$_3$ + H + O$_2$ (Experiment 2.0) Experiments

| peak position, cm$^{-1}$ | peak position, μm | molecule | ref |
|---|---|---|---|
| 851 | 11.75 | HCCH$_2$CH$_3$ | b, c |
| 917 | 10.91 | H$_3$CCH=CH$_2$ and H$_3$CCH$_2$CH$_3$ | b, c,d,e |
| 997 | 10.03 | H$_3$CCH=CH$_2$ | e |
| 1030 | 9.71 | H$_3$CC≡CH | this work |
| 1370 | 7.30 | H$_3$CC≡CH, H$_3$CCH=CH$_2$ and H$_3$CCH$_2$CH$_3$ | this work; b, c,d,e |
| 1384 | 7.23 | H$_3$CC≡CH and H$_3$CCH$_2$CH$_3$ | this work; b, c,d |
| 1439 | 6.95 | H$_3$CC≡CH and H$_3$CCH=CH$_2$ | this work; e |
| 1466 | 6.82 | H$_3$CCH=CH$_2$ and H$_3$CCH$_2$CH$_3$ | b, c,d,e |
| 1644 | 6.08 | H$_3$CCH=CH$_2$ | e |
| 1669 | 5.99 | H$_3$CCH=CHOH/H$_3$CCOH=CH$_2$[a] | f |

[a]Indicates tentative identification. [b]National Institute of Standards and Technology. [c]Reference 82. [d]Reference 83. [e]Reference 81. [f]Reference 84.

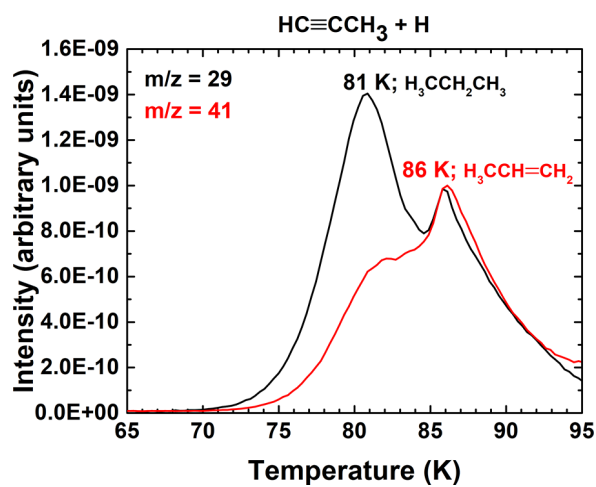

Figure 2. TPD-QMS fragment spectra acquired after deposition of HC≡CCH$_3$ + H (exp 1.2) on a 10 K surface. The main desorption peaks of newly formed H$_3$CCH$_2$CH$_3$ and H$_3$CCH=CH$_2$ are shown at 81 and 86 K, respectively, by $m/z$ = 29 (C$_2$H$_5^+$) and 41 (C$_3$H$_5^+$). A HC≡CCH$_3$:H$_3$CCH=CH$_2$:H$_3$CCH$_2$CH$_3$ abundance ratio of 9:1:2 is measured.

The abundance of H$_3$CCH$_2$CH$_3$ is measured to be 2−3 times greater than that of H$_3$CCH=CH$_2$. This infers that the hydrogenation of H$_3$CCH=CH$_2$ to form H$_3$CCH$_2$CH$_3$ is faster than the hydrogenation of HC≡CCH$_3$ to yield H$_3$CCH=CH$_2$. The work of Kobayashi et al.[64] reported a similar result for the two-carbon equivalents, H$_2$C=CH$_2$ and H$_3$CCH$_3$, where the effective hydrogenation reaction rate constant was found to be ∼3 times higher for H$_3$CCH$_3$ than for H$_2$C=CH$_2$.

**3.2. Inclusion of Hydroxylation (OH) into the HC≡CCH$_3$ Hydrogenation Network.** *3.2.1. Experimental Evidence of n-Propanol and Isopropanol Formation.* As shown in Figure 1, the number of overlapping bands makes it difficult to discern between the RAIR features of newly formed H$_3$CCH=CH$_2$ and H$_3$CCH$_2$CH$_3$, with only two distinct bands apparent. The RAIR spectra become even more convoluted when O$_2$ is added to the mixture. Figure 3 displays the RAIR spectrum of H$_3$CC≡CH + H + O$_2$, in addition to four control RAIR spectra, to attempt characterization of the infrared bands in experiment (exp) 2.0. Spectra of *n*-propanol and isopropanol are compared as they are expected products from the H$_3$CC≡CH + H + O$_2$ experiment and are also commercially available and feasible for UHV conditions. Comparison of the n-propanol and isopropanol spectra to the HC≡CCH$_3$ + H spectrum shows that many of the RAIR features overlap with each other. Additionally, some bands that arise in the H$_3$CC≡CH + H + O$_2$ experiments are difficult to identify (asterisked in Figure 3) as expected, since the hydrogenation of a three-carbon species with O$_2$ is a relatively complex molecular reaction. Thus, the overlapping infrared signals belonging to identified and unidentified molecules make it complicated to track the formation of n-propanol and isopropanol in the infrared at 10 K or even in temperature-dependent RAIR spectra. As discussed in Ioppolo et al.,[87] the formation of complex organic molecules at low temperatures can alternatively be shown by TPD-QMS experiments.

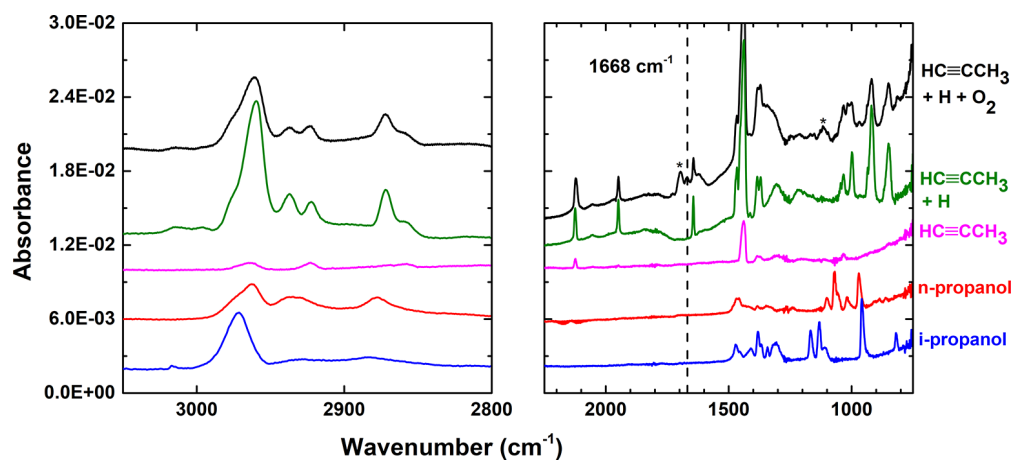

Figure 3. (Top to bottom) RAIR spectra acquired after deposition of HC≡CCH$_3$ + H + O$_2$ (exp 2.0; column density of 1 × 10$^{16}$ cm$^{-2}$), HC≡CCH$_3$ + H (exp 1.1), HC≡CCH$_3$ (exp 1.0), *n*-propanol (exp 3.0; column density of 1 × 10$^{16}$ cm$^{-2}$), and isopropanol (exp 3.1; column density of 1 × 10$^{16}$ cm$^{-2}$) on a 10 K surface. An asterisk (*) indicates peaks that are unidentified, and the band highlighted with a dashed line in the top spectrum is likely due to *n*-propanol or isopropenol. These are further discussed in section 3.2.2. RAIR spectra are offset for clarity.

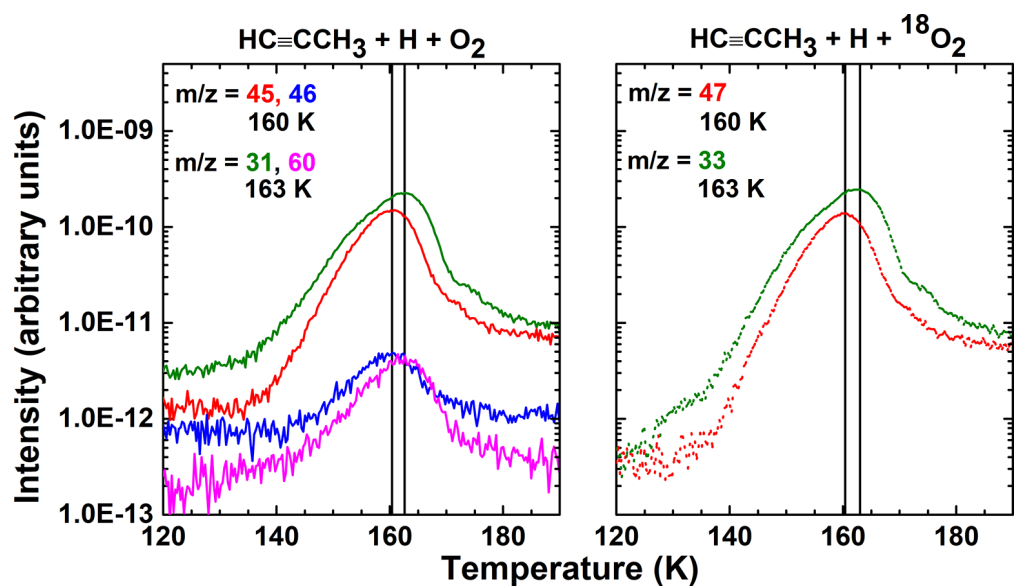

**Figure 4.** TPD-QMS fragment spectra acquired after deposition of HC≡CCH$_3$ + H + O$_2$ (left; exp 2.0) and HC≡CCH$_3$ + H + $^{18}$O$_2$ (right; exp 2.1) on a 10 K surface. The desorption peaks of newly formed isopropanol and n-propanol are shown at 160 and 163 K, respectively. An n-propanol:isopropanol average abundance ratio of 1:1 is measured.

Figure 4 shows the desorption of newly formed n-propanol and isopropanol (and their isotopic counterparts) in two different isotope experiments, where an average abundance ratio of 1:1 for n-propanol:isopropanol is measured. In the H$_3$CC≡CH + H + O$_2$ experiment, the m/z values with the highest intensities for n-propanol and isopropanol are 31 and 45, respectively. Other fragments, such as m/z = 60 and 46, are also shown. In the H$_3$CC≡CH + H + $^{18}$O$_2$ experiment, the m/z values bump up to 33 and 47, respectively, due to isotopically enhanced oxygen. The fragmentation patterns that represent the desorptions of n-propanol and isopropanol are shown in Figure 5. For the desorption of n-propanol at 163 K, the measured relative intensities are 100:2 for m/z = 31:60, 100:3 for m/z = 33:62, and 100:2 for m/z = 31:60. Concerning the desorption of formed isopropanol at 160 K, the relative intensities are also consistent, with ratios of 100:3 for m/z = 45:46, 47:48, and 45:46. The consistency of the relative intensities found between the isotope experiments, as well as between the isotope experiments and the pure n-propanol and isopropanol experiments, further supports the confirmation of solid-state formation of both propanols.

The desorption temperature acts as a further diagnostic, in that the desorption temperature of certain species can shift when they are trapped by relatively less volatile species.[88] An example of this is demonstrated in Supporting Information Figure S1, where the peak desorption of pure isopropanol is seen at 150 K and shifts to 160 K upon addition of H and O$_2$, which is the desorption temperature observed for newly formed isopropanol shown in Figure 4.

*3.2.2. Tentative Experimental Evidence of the Formation of Other Oxygen-Bearing COMs: n-Propenol and Isopropenol.* The H$_3$CC≡CH + H + O$_2$ experiment is expected to generate a variety of COMs that unfortunately not only pushes the limits of the TPD-QMS technique in unambiguously distinguishing the different products formed but also yields species that are not commercially available for control purposes or are challenging to use in a UHV setup. Example reaction products are n-propenol and isopropenol, which are not commercially available as they undergo keto−enol tautomerism at room temperature to primarily form propanal and acetone, respectively.[84] However, at lower temperatures, the enol form becomes more stabilized;[89] thus, n-propenol and isopropenol ices can be present under the applied experimental conditions following addition of OH to the triple bond of H$_3$CC≡CH. Tentative identifications in the RAIR and TPD-QMS data are discussed below. Note that although there is partial evidence for n-propenol and isopropenol formation from the experimental data, the inclusion of computationally derived results confirms their presence in the experiments and is discussed in a later section.

The dashed line in Figure 3 shows the potential identification of n-propenol or isopropenol marked at 1668 cm$^{-1}$ in the H$_3$CC≡CH + H + O$_2$ experiment. This feature is a likely candidate for the C=C stretching mode of propenols. This peak does not overlap with infrared signatures in the H$_3$CC≡CH + H, H$_3$CC≡CH, n-propanol, or isopropanol experiments, meaning it does not represent a product or reactant from those experiments. It also does not red shift in the H$_3$CC≡CH + H + $^{18}$O$_2$ experiment (not shown here); therefore the correlated functional group does not include oxygen. As shown in Shaw et al.,[84] n-propenol and isopropenol have strong absorptions for the C=C stretches at 1684 and 1678 cm$^{-1}$, respectively. Our value of 1668 cm$^{-1}$ is expected, as solid-state infrared frequencies can be red-shifted from that of the gas-phase due to the ice matrix (e.g., CO$_2$).[90] The C=C stretching mode vibrational modes with relatively high band strengths (i.e., the COH bend at ~1100 cm$^{-1}$ and OH stretch at ~3600 cm$^{-1}$)[84] unfortunately overlap with modes of multiple products in the H$_3$CC≡CH + H + O$_2$ experiment. As a direct consequence, an unambiguous identification of propenol in the RAIR data is currently not possible.

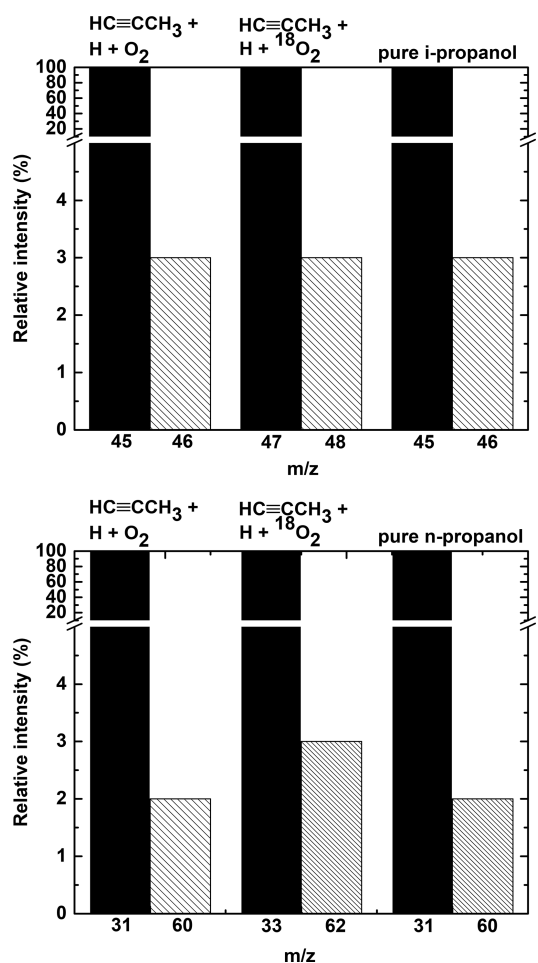

**Figure 5.** QMS fragmentation pattern of two m/z values that are normalized to the QMS signal of the CHOHCH$_3^+$ ion (top) and CH$_2$OH$^+$ ion (bottom) found in the HC≡CCH$_3$ + H + O$_2$ (exp 2.0), HC≡CCH$_3$ + H + $^{18}$O$_2$ (exp 2.1), and pure isopropanol and n-propanol experiments (exps 3.0 and 3.1, respectively) for temperatures of 160 K (top) and 163 K (bottom).

For TPD-QMS, since electron impact ionization fragmentation patterns of n-propenol and isopropenol are not available, simple assumptions have to be used to theoretically derive the possible fragment m/z values. With an electron energy of 70 eV, single bonds can easily break upon dissociative ionization. For both propenols, this results in species with m/z pairs of 43 (C$_2$H$_2$OH$^+$) and 15 (CH$_3^+$), and 41 (C$_3$H$_5^+$) and 17 (OH$^+$). n-Propenol and isopropenol can also remain intact (non-dissociative ionization), which will result in a m/z = 58 signal. Removal of an H atom from the O/C atom of propenol results in a signal for m/z = 57. In the H$_3$CC≡CH + H + $^{18}$O$_2$ experiment, these values bump up to 45 and 15; 41 and 19; 60; and 59, respectively. The signals for some of these m/z values are shown in Figure 6. Two desorption peaks are displayed with peak desorption temperatures of around ~153 and ~159 K in the regular and isotopically enhanced experiments, which is in the range for which the desorption of propenols is expected. From our estimated propenol fragment results, it is not possible to conclude which desorption peak corresponds to which propenol desorption. It should be stressed that no positive identification for the tautomers, propanal and acetone, could be found at their corresponding desorption temperatures of 125 K (Qasim et al., submitted for publication) and 133 K,[91] respectively. The signal of m/z = 43 from ~130 to 170 K has some overlap with the signals of m/z = 57 and 58 and appears to contain other oxygen-containing COMs due to its broad and bumpy desorption profile. Like those of n-propanol and isopropanol, the desorption temperatures of tentatively assigned n-propenol and isopropenol are not that far apart. However, the difference in desorption temperature between isomers also varies depending on the isomers involved. The peak desorption temperatures of around ~153 and ~159 K shown in Figure 6 are within the range of n-propenol desorption of 146−185 K[35] and are also between the peak desorption temperatures of propanal (125 K) (Qasim et al., submitted for publication) and n-propanol (160 K). This is expected when comparing the desorption temperature trend to is also one of the strongest bands of propenol,[84] and thus has the highest probability to be visible in our data. The other propenol that of the two-carbon counterparts, acetaldehyde (H$_3$CCHO), vinyl alcohol (H$_2$CCHOH), and ethanol (H$_3$CCH$_2$OH), which have peak desorption temperatures of 131, 146, and 164 K, respectively (Chuang et al., manuscript in preparation). Yet, without mass spectra and information on the desorption temperatures and profiles of pure n-propenol and isopropenol, the formation of both species in the H$_3$CC≡CH + H + O$_2$ experiment can only be concluded as tentative.

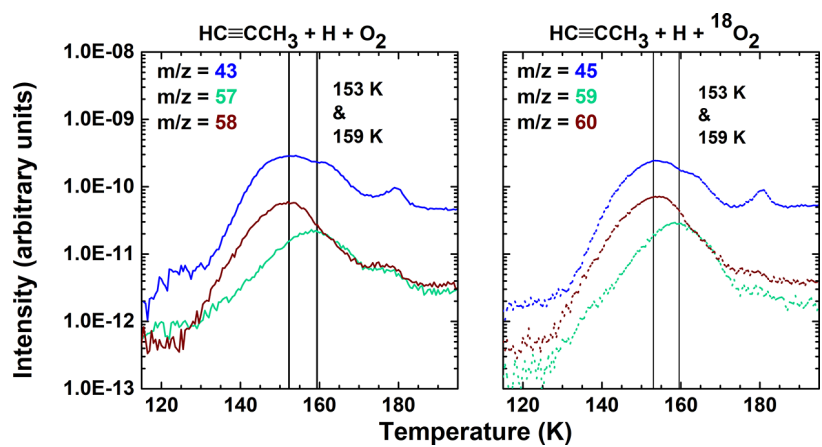

**Figure 6.** TPD-QMS fragment spectra acquired after deposition of HC≡CCH$_3$ + H + O$_2$ (left; exp 2.2) and HC≡CCH$_3$ + H + $^{18}$O$_2$ (right; exp 2.3) on a 10 K surface. n-Propenal and isopropenol are tentatively identified. It cannot be distinguished which of the two propenol isomers desorbs at 153 K and which desorbs at 159 K. The desorption features at ~180 K are due to the co-desorption of both species with H$_2$O$_2$ (an abundant product of H + O$_2$), where both species were trapped in the H$_2$O$_2$ bulk ice.

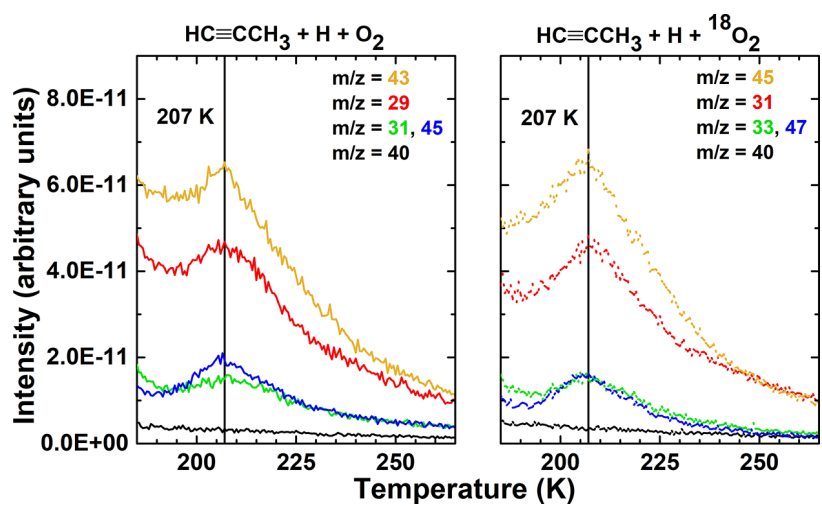

**Figure 7.** TPD-QMS fragment spectra acquired after deposition of HC≡CCH$_3$ + H + O$_2$ (left; exp 2.2) and HC≡CCH$_3$ + H + $^{18}$O$_2$ (right; exp 2.3) on a 10 K surface. Tentatively identified peaks of propane-1,1-diol decomposition product (propanal; $m/z$ = 31, 45), propane-2,2-diol decomposition product (acetone; $m/z$ = 43), and propane-1,2-diol ($m/z$ = 31, 45) are shown. "Blank" refers to the $m/z$ value that does not have a peak desorption in the illustrated temperature range.

*3.2.3. Tentative Experimental Evidence of the Formation of Other Oxygen-Bearing COMs: Three Isomers of Propanediol.* The hydroxylation of n-propenol and isopropenol can lead to the formation of COMs with two oxygens, such as propane-1,1-diol, propane-2,2-diol, and propane-1,2-diol. However, study of the pure samples is difficult due to their chemical instability and low vapor pressure under standard temperature and pressure conditions. Particularly, propane-1,1-diol and propane-2,2-diols are very unstable and are therefore not commercially available. Upon desorption into the gas-phase, the geminal-diol equilibrium of propane-1,1-diol and propane-2,2-diol shifts greatly toward formation of the spontaneous decomposition products, propanal with H$_2$O and acetone with H$_2$O, respectively.[92] Such chemical transformations can be used to tag the formation of propanediol isomers. If the geminal diols are formed in the solid-state, then upon their desorption into the gas-phase following spontaneous decomposition, QMS signatures of propanal, acetone, and H$_2$O would be observed at their noncharacteristic desorption temperature of around 200 K, where propanal and acetone have characteristic peak desorption temperatures of 125 K (Qasim et al., submitted for publication) and 133 K,[91] respectively. As shown in Figure 7, $m/z$ = 43, 29, 31, and 45 represent the main $m/z$ signals of acetone, propanal, and propane-1,2-diol, respectively, according to the NIST database. The peak intensity of these values is found around ∼207 K, which is in line with the temperature of 203 K that was tentatively assigned for propane-1,2-diol and propane-1,3-diol desorption.[93] The expected $m/z$ shifts are also found in the H$_3$CC≡CH + H + $^{18}$O$_2$ experiment, as shown in the right panel of Figure 7. Yet, without confirmation of the desorption temperatures, decomposition products, and fragmentation patterns of the pure samples, only a tentative identification of the isomers of propanediol is reported here.

We summarize all of the experimental results presented here. As shown in section 3.1, H$_3$CCH=CH$_2$ and H$_3$CCH$_2$CH$_3$ are experimentally confirmed to form from the hydrogenation of H$_3$CC≡CH, with a HC≡CCH$_3$:H$_3$CCH= CH$_2$:H$_3$CCH$_2$CH$_3$ abundance ratio of 9:1:2. In section 3.2, it is shown that inclusion of OH leads to the formation of n-propanol and isopropanol, and the average abundance ratio of n-propanol:isopropanol is 1:1. Tentative detections of n-propenol, isopropenol, propane-1,1-diol, propane-2,2-diol, and propane-1,2-diol are found. The tautomers of n-propenol (propanal) and isopropenol (acetone), as well as the acidic derivative propanoic acid, are not detected.

## 4. ENERGIES AND FORMATION MECHANISMS

The experimental findings are joined by computationally derived energy barriers to draw the exact products formed and the correlated formation mechanisms. Computationally derived energies for eight different reactions that occur in the experiments are found in Table 4. The first column of the table lists three types of energies studied: interaction, activation, and reaction. The interaction energy is the energy gained when the pre-reactive complex (PRC) is formed. This complex is formed when the two reactants have enough time or energy to rearrange themselves in the ice before product formation. This is opposite to the situation of the separated reactants (SRs), where the reactants immediately react to form the end product. The activation energy is with respect to both SR and with respect to the PRC, and the difference between these two values is equal to the interaction energy of the PRC. Finally, the reaction energy is defined by the exothermicity.

The main findings from Table 4 are discussed as follows. Comparing the reactivity toward H and OH, in all but one case the reaction of an unsaturated species with an OH radical is more favorable than with an H atom. However, it is important to keep in mind that the reaction with OH results in the formation of a C–O bond. This, contrary to the formation of a C–H bond, is not accelerated much by taking tunneling into account at low temperature. Furthermore, in accordance with results from Kobayashi et al.[64] and Zaverkin et al.,[94] we find that reaction with a double-bonded (C=C) species is easier than with a triple-bonded (C≡C) molecule. It is also confirmed that the exterior carbon is more reactive toward H than the interior carbon.

Table 4. Interaction, Activation, and Reaction Energies for $H_3CC\equiv CH + OH$, $H_3CCH=CH_2 + OH$, $H_3CC\equiv CH + H$, and $H_3CCH=CH_2 + H$ Calculated at the MPWB1K/def2-TZVP Level of theory[a]

| energy type | E or I carbon | SR or PRC | PES | ZPE | Total |
|---|---|---|---|---|---|
| | | $HC\equiv CCH_3 + OH$ | | | |
| interaction | I | | −1342 | 591 | −751 |
| interaction | E | | −2073 | 690 | −1382 |
| activation | I | SR | −126 | 763 | 636 |
| activation | E | SR | −358 | 738 | 380 |
| activation | I | PRC | 1216 | 172 | 1387 |
| activation | E | PRC | 1715 | 48 | 1762 |
| reaction | I | | −16928 | 2089 | −14839 |
| reaction | E | | −18122 | 2333 | −15790 |
| | | $H_2C=CHCH_3 + OH$ | | | |
| interaction | I | | −1543 | 806 | −738 |
| interaction | E | | −1624 | 732 | −892 |
| activation | I | SR | −1490 | 799 | −691 |
| activation | E | SR | −1358 | 687 | −671 |
| activation | I | PRC | 53 | −7 | 47 |
| activation | E | PRC | 266 | −45 | 221 |
| reaction | I | | −15736 | 1692 | −14045 |
| reaction | E | | −16088 | 1825 | −14263 |
| | | $HC\equiv CCH_3 + H$ | | | |
| interaction | I | | −30 | 262 | 232 |
| interaction | E | | −26 | 177 | 151 |
| activation | I | SR | 2378 | 488 | 2866 |
| activation | E | SR | 1447 | 295 | 1742 |
| activation | I | PRC | 2408 | 226 | 2634 |
| activation | E | PRC | 1473 | 118 | 1591 |
| reaction | I | | −19772 | 3207 | −16565 |
| reaction | E | | −22049 | 3123 | −18926 |
| | | $H_2C=CHCH_3 + H$ | | | |
| interaction | I | | −25 | 201 | 176 |
| interaction | E | | −17 | 135 | 118 |
| activation | I | SR | 1614 | 522 | 2136 |
| activation | E | SR | 748 | 361 | 1109 |
| activation | I | PRC | 1639 | 321 | 1960 |
| activation | E | PRC | 765 | 226 | 991 |
| reaction | I | | −19334 | 2717 | −16617 |
| reaction | E | | −21249 | 2660 | −18589 |

[a]E and I refer to the exterior and interior carbon, respectively. SR and PRC refer to the separated reactants and pre-reactive complex, respectively. All values are in units of kelvin.

Whether the reactions proceed with or without a PRC is evaluated from the results from Table 4. When the activation energies for the reactions $H_3CC\equiv CH + OH$ and $H_3CCH=CH_2 + OH$ (where OH is derived from $H + O_2$ in the experiments) are considered with respect to the separated reactants, it is obvious that the reactions should be able to take place very easily as the reaction is either barrierless ($H_3CCH=CH_2 + OH$) or has a relatively low activation energy ($H_3CC\equiv CH + OH$, 380 or 636 K). However, when the two reactants form a PRC, it is expected that the excess energy of the complex formation is dissipated into the ice mantle well before the reaction itself is attempted, as the energy dissipation in ices seems to take place on a picosecond time scale.[95,96] Therefore, the effective activation energy to be overcome increases by the same amount of energy that is gained from the interaction of OH with $H_3CC\equiv CH/H_3CCH=CH_2$ (i.e., the interaction energy is added to determine the total activation energy). This significantly increases the activation energy, although, for the reaction with $H_3CCH=CH_2$, it remains close to barrierless. Which of the two surface "mechanisms" is the best description for these reactions in an ice (be it in the laboratory or in the interstellar medium) can be debated, and it is quite likely that a variety of geometries exist that may lead to an averaged-out effect. For example, if OH is a neighboring species to $H_3CC\equiv CH/H_3CCH=CH_2$, the immediate surroundings may cause steric hindrance between the two species and, therefore, mitigate a favorable orientation of the two species with respect to each other. On the other hand, OH could potentially use its excess energy to rearrange the position and thus obtain a more favorable orientation. Note that for the reaction of H with any molecule, the low diffusion barrier of the H atom always allows for a mechanism that considers the PRC to take place. Therefore, the effect of the relative geometries on the reaction efficiency is less pronounced for species that have low diffusion barriers or high activation barriers.

The likelihood of constitutional isomerization (or tautomerization) for products formed in the experiments is assessed from the computationally derived results in Table 5. From the results, it is immediately clear that direct isomerization reactions cannot take place efficiently in ices in the dense atomic/molecular medium, as the typical activation energy is more than 25,000 K. If the reaction were to be actively catalyzed by another molecule that can simultaneously donate and accept an H atom, such as $H_2O$ or the OH group of $CH_3OH$, the activation energy may drop considerably.[97,98] The value could then drop close to the values involved for H-hopping from an OH group to an OH radical, which is >1800 K for the reaction $CH_3OH + OH$,[99] and between 2500 and 9000 K for the reaction $OH + (H_2O)_n$ ($n = 1-3$).[100] This can only happen, however, if a suitable multispecies geometry can be established in the ice (i.e., if solvation is present).

Table 5. Activation Barriers of the Isomerization Reactions of $n$-Propenol and Isopropenol[a]

| reaction | PES | ZPE | total |
|---|---|---|---|
| acetone → isopropenol | 34452 | −1692 | 32760 |
| isopropenol → acetone | 28686 | −2116 | 26569 |
| propanal → $n$-propenol | 35132 | −1892 | 33240 |
| $n$-propenol → propanal | 31209 | −2100 | 29109 |

[a]All values are in units of kelvin.

The products formed and their formation mechanisms are finally discussed below. It is apparent from Table 4 that the activation barriers of $H_3CC\equiv CH + OH$ are primarily lower than those of $H_3CC\equiv CH + H$ and by a substantial amount. This is even more pronounced when comparing the activation barriers of $H_3CCH=CH_2 + OH$ and $H_3CCH=CH_2 + H$. The formation of propanols in the experiments indicates that OH addition is relatively efficient under our experimental conditions. However, H atoms are more mobile and can effectively tunnel at low temperatures, making H addition competitive to OH addition. Therefore, from the combination of computationally derived activation barriers and the unambiguous identification of propanol formation, it is found that $H_3CC\equiv CH$ is effectively attacked by OH radicals more so than by H atoms only when both species neighbor $H_3CC\equiv CH$ and when OH is oriented in a favorable position for reaction. According to Table 4, this narrows the selection of activation energies to four values: 636 and 1387 K (OH attack on the interior carbon) and 380 and 1762 K (OH attack on the exterior carbon). The experimental results give an $n$-propanol:isopropanol average abundance ratio of 1:1, depending on which $m/z$ values are used to determine the relative abundance. Therefore, it is likely that both species are formed with comparable abundances (i.e., attack to the exterior and interior carbons occurs equally under our experimental conditions). Moreover, this conclusion is fully in line with the idea that, for reactants that can strongly interact with each other, the efficiency of both "mechanisms" is influenced by the geometric orientation of the formed OH radical with respect to the $H_3CC\equiv CH$ molecule (i.e., both "mechanisms" are actively participating). If this was not the case, then it is expected that $n$-propanol would be distinctly more abundant in the experiments, as the lowest activation barrier of 380 K would favor $n$-propanol formation. The outcome of the experimental and theoretical results pieced together leads to the formation mechanisms that are most likely occurring in the $H_3CC\equiv CH + H + O_2$ experiment, which are illustrated in Figure 8. The activation barriers for some of the displayed reaction steps are listed in Table 4. As shown, the formation pathways of $n$-propanol and isopropanol include the formation of $n$-propenol and isopropenol. Thus, $n$-propenol and isopropenol are products formed in our experiments. Since the activation barriers of OH addition to propenols are not investigated in this work, the propanediol isomers remain to be tentative reaction products, as noted in Figure 8. For the $H_3CC\equiv CH + H$ experiment, the relatively high barriers show that $H_3CC\equiv CH$ and H do not have a strong interaction with each other such as that of $H_3CC\equiv CH$ and OH. Therefore, the geometric orientation of the two species has less of an effect on which "mechanism" would proceed. H attack on the exterior carbon of $H_3CC\equiv CH$ and $H_3CCH=CH_2$ results in the lower activation barrier in comparison to H attack on the interior carbon, and thus is proposed to be the more likely scenario involved in $H_3CCH=CH_2$ and $H_3CCH_2CH_3$ formation.

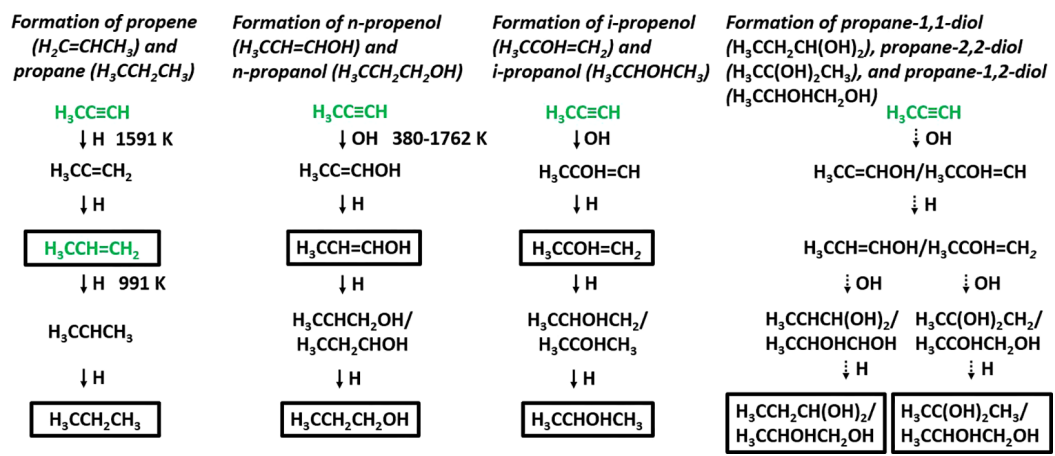

**Figure 8.** Proposed mechanisms for exps 2.0 and 2.2. Relevant species within each mechanism are boxed. Species labeled with green font are those that have been detected in space. Dotted boxes indicate tentatively identified species in this study.

## 5. ASTROCHEMICAL AND ASTROBIOLOGICAL IMPLICATIONS

The formation of three-carbon chains, alcohols, and, to an extent, geminal diols from $H_3CC\equiv CH + H/OH$ at the low temperature of 10 K can take place at the interface of carbonaceous grains and $H_2O$-rich ice in the dark cloud stage of stellar formation. At an $A_V < 1.5$, $H_2O$ is just starting to coat interstellar grains partially by the accretion of H and O atoms.[32] These atoms combine on the grain surface to form OH radicals and have the potential to react with the carbon-rich constituents that are at the surface of the dust grain. As the visual extinction grows to 3, a $H_2O$-rich ice is formed on top of the mineral−ice interface. The icy grain is further coated by other molecules (e.g., CO) as it travels through various molecular freeze-out stages.[32]

The formation of $n$-propanol and isopropanol and propenol shown in this combined experimental and theoretical study brings to light possible formation pathways for these species in astrochemical environments. Particularly, the formation of solid-state isopropanol and propenol is intriguing, as there are no experimentally based studies on these molecules to our knowledge. It has been shown by Qasim et al. (submitted for publicataion) and Abplanalp et al.[35] that $n$-propanol and propenol can be formed in a CO-rich ice, respectively. For the first time experimentally, it is shown here that all four alcohols can be formed before the heavy CO freeze-out stage (i.e., in the $H_2O$-rich ice phase) and along the same formation route. Thus, astronomical surveys may be able to identify all four species simultaneously.

This route can be extrapolated to polyynes containing $H_3C-(C\equiv C)_n-H$ structures and therefore have an astrobiological context. As a $H_3CC\equiv CH$-containing ice yields $n$-propanol, similarly ices with $H_3C-(C\equiv C)_n-H$ structures can lead to the formation of fatty alcohols (i.e., long-chain alcohols), which are found to be constituents of simple lipids. For example, the incorporation of dodecanol to form primitive-like lipid bilayer membranes has been shown.[101] Moreover waxes, which are simple lipids, are composed of fatty alcohols which fatty acids are connected with. This includes the wax Spermaceti, which contains hexadecan-1-ol ($H_3C(CH_2)_{14}CH_2OH$), and beeswax and carnuba, which both contain triacontan-1-ol ($H_3C-(CH_2)_{28}CH_2OH$).[102]

The hydroxylation of polyynes discussed in our study provides a valid mechanism for the formation of various linear alcohols starting from carbon chains. Such alcohols may be present in the bottom layer (layer closest to the grain surface) of $H_2O$-rich interstellar ices. Such layering is advantageous to the preservation of such interface reaction products, in that bulk $H_2O$ ice can partially block UV light.[103,104] From there, the possibility increases for such prebiotic material to be safely transferred to the early Earth and contribute to the formation of primitive cell membranes.

## 6. CONCLUSIONS

This combined experimental and computational study provides insights into what and how solid-state alcohols may be formed in the $H_2O$-rich ice phase of cold molecular cores. The main conclusions from the conjoined results are highlighted below:

(1) $H_3CC\equiv CH$ (propyne) + H efficiently forms $H_3CCH=CH_2$ (propene) and $H_3CCH_2CH_3$ (propane), where the abundance ratio of the three species is 9:1:2, respectively. The experimental result is in line with the computational results, in that the barrier to hydrogenate C=C is lower than to hydrogenate C≡C.

(2) The experimental investigation shows that $H_3CC\equiv CH$ + OH leads to the formation of n-propanol and isopropanol at 10 K under "non-energetic" (without UV, cosmic rays, and so on and/or other "energetic" particles) conditions. The formation of n-propenol and isopropenol in our experiments is confirmed from the combination of the experimental and theoretical results. Tentative identifications of propane-1,1-diol, propane-2,2-diol, and propane-1,2-diol are found from the experimental data.

(3) The formation yield of n-propanol (and thus n-propenol) in the experiments is observed to be comparable to that of isopropanol (and thus isopropenol), with an experimentally measured n-propanol:isopropanol average abundance ratio of 1:1. This value is in line with the computational calculations and the finding that both "mechanisms" (PRC and SR) are equally influential due to the prominent role of how $H_3CC\equiv CH$ and OH are oriented toward each other.

(4) OH addition to $H_3CC\equiv CH$ is observed to be more effective than H addition when both radicals are in close vicinity to $H_3CC\equiv CH$ and when the OH radical is situated in a favorable orientation for reaction. This is supported by the relatively low computationally derived activation barriers of $H_3CC\equiv CH$ + OH and the formation of propanols in the experiments.

(5) Propanols and propenols, and to an extent propanediols, are expected to form simultaneously in suitable ISM regions and may have an icy origin at the very beginning of the dark cloud stage.

(6) The presented formation routes may be extended to polyynes with $H_3C-(C\equiv C)_n-H$ structures. These structures can transform into fatty alcohols, which are the components of simple lipids which primitive cell membranes were likely, in part, assembled by.

## ASSOCIATED CONTENT

### Supporting Information

The Supporting Information is available free of charge on the ACS Publications website at DOI: 10.1021/acsearthspacechem.9b00062.

> Additional TPD-QMS spectra (Figure S1), pathways and benchmark calculations, IRC paths (Figures S2 and S3), NEB paths (Figures S4 and S5), M06-2X validity check (Table S1), and CCSD(T)-F12/cc-VDZ-F12 validity check (Table S2) (PDF)

## AUTHOR INFORMATION

### Corresponding Author
*E-mail: dqasim@strw.leidenuniv.nl.

### ORCID
Danna Qasim: 0000-0002-3276-4780
Thanja Lamberts: 0000-0001-6705-2022
Jiao He: 0000-0003-2382-083X
Johannes Kästner: 0000-0001-6178-7669
Harold Linnartz: 0000-0002-8322-3538

### Present Address
#Laboratory Astrophysics Group of the Max Planck Institute for Astronomy at the Friedrich Schiller University Jena, Institute of Solid State Physics, Helmholtzweg 3, D-07743 Jena, Germany.

### Notes
The authors declare no competing financial interest.

## ACKNOWLEDGMENTS

This research is financially supported by the Dutch Astrochemistry Network II (DANII). Further support includes a VICI grant of NWO (The Netherlands Organization for Scientific Research) and funding by NOVA (The Netherlands Research School for Astronomy). T.L. is supported by NWO via a VENI fellowship (722.017.00). G.F. acknowledges financial support from the European Union's Horizon 2020 research and innovation programme under the Marie Sklodowska-Curie Grant Agreement No. 664931. S.I. thanks the Royal Society for financial support and the Holland Research School for Molecular Chemistry (HRSMC) for a travel grant.

## REFERENCES

(1) Cherchneff, I. The formation of polycyclic aromatic hydrocarbons in evolved circumstellar environments. *EAS Publ. Ser.* **2011**, *46*, 177−189.

(2) Contreras, C. S.; Salama, F. Laboratory investigations of polycyclic aromatic hydrocarbon formation and destruction in the circumstellar outflows of carbon stars. *Astrophys. J., Suppl. Ser.* **2013**, *208*, 6.

(3) Tielens, A. The molecular universe. *Rev. Mod. Phys.* **2013**, *85*, 1021−1081.

(4) Castellanos, N. P. *Breaking & Entering: PAH photodissociation and top-down chemistry*. Ph.D. thesis, Leiden University, Leiden, The Netherlands, 2018.

(5) Woods, P. M.; Millar, T. J.; Herbst, E.; Zijlstra, A. A. The chemistry of protoplanetary nebulae. *Astron. Astrophys.* **2003**, *402*, 189−199.

(6) Cuylle, S. H.; Zhao, D.; Strazzulla, G.; Linnartz, H. Vacuum ultraviolet photochemistry of solid acetylene: A multispectral approach. *Astron. Astrophys.* **2014**, *570*, A83.

(7) Pascoli, G.; Polleux, A. Condensation and growth of hydrogenated carbon clusters in carbon-rich stars. *Astron. Astrophys.* **2000**, *359*, 799−810.

(8) Snyder, L.; Buhl, D. Interstellar methylacetylene and isocyanic acid. *Nature, Phys. Sci.* **1973**, *243*, 45−46.

(9) Irvine, W.; Hoglund, B.; Friberg, P.; Askne, J.; Ellder, J. The increasing chemical complexity of the Taurus dark clouds-Detection of CH3CCH and C4H. *Astrophys. J.* **1981**, *248*, L113−L117.

(10) Kuiper, T.; Kuiper, E. R.; Dickinson, D. F.; Turner, B.; Zuckerman, B. Methyl acetylene as a temperature probe for dense interstellar clouds. *Astrophys. J.* **1984**, *276*, 211−220.

(11) Cernicharo, J.; Heras, A. M.; Pardo, J. R.; Tielens, A.; Guélin, M.; Dartois, E.; Neri, R.; Waters, L. Methylpolyynes and small hydrocarbons in CRL 618. *Astrophys. J.* **2001**, *546*, L127−L130.

(12) Kaifu, N.; Ohishi, M.; Kawaguchi, K.; Saito, S.; Yamamoto, S.; Miyaji, T.; Miyazawa, K.; Ishikawa, S.-i.; Noumaru, C.; Harasawa, S.; Okuda, M.; Suzuki, H. A 8.8−50 GHz complete spectral line survey toward TMC-1 I. Survey data. *Publ. Astron. Soc. Jpn.* **2004**, *56*, 69−173.

(13) Agúndez, M.; Fonfría, J. P.; Cernicharo, J.; Pardo, J.; Guélin, M. Detection of circumstellar CH2CHCN, CH2CN, CH3CCH, and H2CS. *Astron. Astrophys.* **2008**, *479*, 493−501.

(14) Muller, S.; Beelen, A.; Guélin, M.; Aalto, S.; Black, J. H.; Combes, F.; Curran, S.; Theule, P.; Longmore, S. Molecules at z= 0.89-A 4-mm-rest-frame absorption-line survey toward PKS 1830- 211. *Astron. Astrophys.* **2011**, *535*, A103.

(15) Malek, S. E.; Cami, J.; Bernard-Salas, J. The rich circumstellar chemistry of SMP LMC 11. *Astrophys. J.* **2012**, *744*, 16.

(16) Qiu, J.; Wang, J.; Shi, Y.; Zhang, J.; Fang, M.; Li, F. Deep millimeter spectroscopy observations toward NGC 1068. *Astron. Astrophys.* **2018**, *613*, A3.

(17) Herbst, E.; van Dishoeck, E. F. Complex organic interstellar molecules. *Annu. Rev. Astron. Astrophys.* **2009**, *47*, 427−480.

(18) Miyauchi, N.; Hidaka, H.; Chigai, T.; Nagaoka, A.; Watanabe, N.; Kouchi, A. Formation of hydrogen peroxide and water from the reaction of cold hydrogen atoms with solid oxygen at 10 K. *Chem. Phys. Lett.* **2008**, *456*, 27−30.

(19) Ioppolo, S.; Cuppen, H.; Romanzin, C.; van Dishoeck, E.; Linnartz, H. Laboratory evidence for efficient water formation in interstellar ices. *Astrophys. J.* **2008**, *686*, 1474−1479.

(20) Ioppolo, S.; Cuppen, H.; Romanzin, C.; van Dishoeck, E.; Linnartz, H. Water formation at low temperatures by surface $O_2$ hydrogenation I: characterization of ice penetration. *Phys. Chem. Chem. Phys.* **2010**, *12*, 12065−12076.

(21) Matar, E.; Congiu, E.; Dulieu, F.; Momeni, A.; Lemaire, J. Mobility of D atoms on porous amorphous water ice surfaces under interstellar conditions. *Astron. Astrophys.* **2008**, *492*, L17−L20.

(22) Cuppen, H.; Ioppolo, S.; Romanzin, C.; Linnartz, H. Water formation at low temperatures by surface $O_2$ hydrogenation II: the reaction network. *Phys. Chem. Chem. Phys.* **2010**, *12*, 12077−12088.

(23) Romanzin, C.; Ioppolo, S.; Cuppen, H.; van Dishoeck, E.; Linnartz, H. Water formation by surface O3 hydrogenation. *J. Chem. Phys.* **2011**, *134*, 084504.

(24) van Dishoeck, E. F.; Herbst, E.; Neufeld, D. A. Interstellar water chemistry: from laboratory to observations. *Chem. Rev.* **2013**, *113*, 9043−9085.

(25) Hama, T.; Watanabe, N. Surface processes on interstellar amorphous solid water: adsorption, diffusion, tunneling reactions, and nuclear-spin conversion. *Chem. Rev.* **2013**, *113*, 8783−8839.

(26) Chang, Q.; Herbst, E. A unified microscopic-macroscopic Monte Carlo simulation of gas-grain chemistry in cold dense interstellar clouds. *Astrophys. J.* **2012**, *759*, 147.

(27) Chyba, C. F.; Thomas, P. J.; Brookshaw, L.; Sagan, C. Cometary delivery of organic molecules to the early Earth. *Science* **1990**, *249*, 366−373.

(28) Moran, L. A.; Horton, H. R.; Scrimgeour, G.; Perry, M. *Principles of biochemistry*; Pearson: Upper Saddle River, NJ, USA, 2012.

(29) Deamer, D.; Dworkin, J. P.; Sandford, S. A.; Bernstein, M. P.; Allamandola, L. J. The first cell membranes. *Astrobiology* **2002**, *2*, 371−381.

(30) Budin, I.; Szostak, J. W. Physical effects underlying the transition from primitive to modern cell membranes. *Proc. Natl. Acad. Sci. U. S. A.* **2011**, *108*, 5249−5254.

(31) De Rosa, M.; Gambacorta, A.; Gliozzi, A. Structure, biosynthesis, and physicochemical properties of archaebacterial lipids. *Microbiol. Mol. Biol. Rev.* **1986**, *50*, 70−80.

(32) Boogert, A.; Gerakines, P. A.; Whittet, D. C. Observations of the icy universe. *Annu. Rev. Astron. Astrophys.* **2015**, *53*, 541−581.

(33) Bernstein, M. P.; Sandford, S. A.; Allamandola, L. J.; Chang, S.; Scharberg, M. A. Organic compounds produced by photolysis of realistic interstellar and cometary ice analogs containing methanol. *Astrophys. J.* **1995**, *454*, 327−344.

(34) Chen, Y.-J.; Ciaravella, A.; Munoz Caro, G.; Cecchi-Pestellini, C.; Jiménez-Escobar, A.; Juang, K.-J.; Yih, T.-S. Soft X-Ray irradiation of methanol ice: formation of products as a function of photon energy. *Astrophys. J.* **2013**, *778*, 162.

(35) Abplanalp, M. J.; Gozem, S.; Krylov, A. I.; Shingledecker, C. N.; Herbst, E.; Kaiser, R. I. A study of interstellar aldehydes and enols as tracers of a cosmic ray-driven nonequilibrium synthesis of complex organic molecules. *Proc. Natl. Acad. Sci. U. S. A.* **2016**, *113*, 7727−7732.

(36) Paardekooper, D.; Bossa, J.-B.; Linnartz, H. Laser desorption time-of-flight mass spectrometry of vacuum UV photo-processed methanol ice. *Astron. Astrophys.* **2016**, *592*, A67.

(37) Fedoseev, G.; Cuppen, H. M.; Ioppolo, S.; Lamberts, T.; Linnartz, H. Experimental evidence for glycolaldehyde and ethylene glycol formation by surface hydrogenation of CO molecules under dense molecular cloud conditions. *Mon. Not. R. Astron. Soc.* **2015**, *448*, 1288−1297.

(38) Butscher, T.; Duvernay, F.; Theule, P.; Danger, G.; Carissan, Y.; Hagebaum-Reignier, D.; Chiavassa, T. Formation mechanism of glycolaldehyde and ethylene glycol in astrophysical ices from HCO and CH2OH recombination: an experimental study. *Mon. Not. R. Astron. Soc.* **2015**, *453*, 1587−1596.

(39) Chuang, K.-J.; Fedoseev, G.; Ioppolo, S.; van Dishoeck, E.; Linnartz, H. H-atom addition and abstraction reactions in mixed CO, H 2 CO and CH 3 OH ices−an extended view on complex organic molecule formation. *Mon. Not. R. Astron. Soc.* **2016**, *455*, 1702−1712.

(40) Butscher, T.; Duvernay, F.; Rimola, A.; Segado-Centellas, M.; Chiavassa, T. Radical recombination in interstellar ices, a not so simple mechanism. *Phys. Chem. Chem. Phys.* **2017**, *19*, 2857−2866.

(41) Fuchs, G.; Cuppen, H.; Ioppolo, S.; Romanzin, C.; Bisschop, S.; Andersson, S.; van Dishoeck, E.; Linnartz, H. Hydrogenation reactions in interstellar CO ice analogues-A combined experimental/theoretical approach. *Astron. Astrophys.* **2009**, *505*, 629−639.

(42) Ball, J. A.; Gottlieb, C. A.; Lilley, A.; Radford, H. Detection of methyl alcohol in sagittarius. *Astrophys. J.* **1970**, *162*, L203−L210.

(43) Zuckerman, B.; Turner, B.; Johnson, D.; Clark, F.; Lovas, F.; Fourikis, N.; Palmer, P.; Morris, M.; Lilley, A.; Ball, J.; Gottlieb, C. A.; Litvak, M. M.; Penfield, H. Detection of interstellar trans-ethyl alcohol. *Astrophys. J.* **1975**, *196*, L99−L102.

(44) Turner, B. E.; Apponi, A. J. Microwave detection of interstellar vinyl alcohol, CH2= CHOH. *Astrophys. J.* **2001**, *561*, L207−L210.

(45) Hollis, J. M.; Lovas, F. J.; Jewell, P. R.; Coudert, L. Interstellar antifreeze: ethylene glycol. *Astrophys. J.* **2002**, *571*, L59−L62.

(46) McGuire, B. A.; Shingledecker, C. N.; Willis, E. R.; Burkhardt, A. M.; El-Abd, S.; Motiyenko, R. A.; Brogan, C. L.; Hunter, T. R.; Margulès, L.; Guillemin, J.-C.; Garrod, R. T.; Herbst, E.; Remijan, A. J. ALMA detection of interstellar methoxymethanol (CH3OCH2OH). *Astrophys. J., Lett.* **2017**, *851*, L46.

(47) Boogert, A.; Huard, T.; Cook, A.; Chiar, J.; Knez, C.; Decin, L.; Blake, G.; Tielens, A.; van Dishoeck, E. Ice and dust in the quiescent medium of isolated dense cores. *Astrophys. J.* **2011**, *729*, 92.

(48) van't Hoff, M. L.; Tobin, J. J.; Trapman, L.; Harsono, D.; Sheehan, P. D.; Fischer, W. J.; Megeath, S. T.; van Dishoeck, E. F. Methanol and its relation to the water snowline in the disk around the young outbursting star V883 Ori. *Astrophys. J., Lett.* **2018**, *864*, L23.

(49) Ioppolo, S.; Fedoseev, G.; Lamberts, T.; Romanzin, C.; Linnartz, H. SURFRESIDE2: An ultrahigh vacuum system for the investigation of surface reaction routes of interstellar interest. *Rev. Sci. Instrum.* **2013**, *84*, 073112.

(50) Qasim, D.; Chuang, K.-J.; Fedoseev, G.; Ioppolo, S.; Boogert, A.; Linnartz, H. Formation of interstellar methanol ice prior to the heavy CO freeze-out stage. *Astron. Astrophys.* **2018**, *612*, A83.

(51) Chuang, K.-J.; Fedoseev, G.; Qasim, D.; Ioppolo, S.; van Dishoeck, E.; Linnartz, H. H2 chemistry in interstellar ices: the case of CO ice hydrogenation in UV irradiated CO: H2 ice mixtures. *Astron. Astrophys.* **2018**, *617*, A87.

(52) Tschersich, K.; Von Bonin, V. Formation of an atomic hydrogen beam by a hot capillary. *J. Appl. Phys.* **1998**, *84*, 4065−4070.

(53) Tschersich, K. Intensity of a source of atomic hydrogen based on a hot capillary. *J. Appl. Phys.* **2000**, *87*, 2565−2573.

(54) Tschersich, K.; Fleischhauer, J.; Schuler, H. Design and characterization of a thermal hydrogen atom source. *J. Appl. Phys.* **2008**, *104*, 034908.

(55) Watanabe, N.; Kouchi, A. Efficient formation of formaldehyde and methanol by the addition of hydrogen atoms to CO in H2O-CO ice at 10 K. *Astrophys. J.* **2002**, *571*, L173−L176.

(56) Watanabe, N.; Shiraki, T.; Kouchi, A. The dependence of H2CO and CH3OH formation on the temperature and thickness of H2O-CO

ice during the successive hydrogenation of CO. *Astrophys. J.* **2003**, *588*, L121−L124.

(57) Cuppen, H.; Herbst, E. Simulation of the formation and morphology of ice mantles on interstellar grains. *Astrophys. J.* **2007**, *668*, 294−309.

(58) Shimanouchi, T. Molecular Vibrational Frequencies. In *NIST Chemistry WebBook*, NIST Standard Reference Database Number 69; Linstrom, P. J., Mallard, W. G., Eds.; National Institute of Standards and Technology: Gaithersburg, MD, USA; https://doi.org/10.18434/T4D303 (retrieved Dec. 4, 2018).

(59) NIST Mass Spec Data Center; Stein, S. E., Director Mass Spectra. In *NIST Chemistry WebBook*, NIST Standard Reference Database Number 69; Linstrom, P. J., Mallard, W. G., Eds.; National Institute of Standards and Technology: Gaithersburg, MD, USA; https://doi.org/10.18434/T4D303 (retrieved Dec. 4, 2018)

(60) Martín-Doménech, R.; Manzano-Santamaría, J.; Munoz Caro, G.; Cruz-Díaz, G. A.; Chen, Y.-J.; Herrero, V. J.; Tanarro, I. UV photoprocessing of $CO_2$ ice: a complete quantification of photochemistry and photon-induced desorption processes. *Astron. Astrophys.* **2015**, *584*, A14.

(61) Kim, Y.-K.; Irikura, K. K.; Rudd, M. E.; Ali, M. A.; Stone, P. M.; Chang, J.; Coursey, J. S.; Dragoset, R. A.; Kishore, A. R.; Olsen, K. J.; Sansonetti, A. M.; Wiersma, G. G.; Zucker, D. S.; Zucker, M. A. *Electron-Impact Ionization Cross Section for Ionization and Excitation Database*, version 3.0 [Online]; National Institute of Standards and Technology: Gaithersburg, MD, USA, 2004; http://physics.nist.gov/ionxsec [accessed Mar. 16, 2019].

(62) Chuang, K. *formation of complex organic molecules in dense clouds: sweet results from the laboratory*. Ph.D. thesis, Leiden University, Leiden, The Netherlands, 2018.

(63) Kolasinski, K. W. *Surface science: Foundations of catalysis and nanoscience*; John Wiley & Sons: West Chester, PA, USA, 2012.

(64) Kobayashi, H.; Hidaka, H.; Lamberts, T.; Hama, T.; Kawakita, H.; Kästner, J.; Watanabe, N. Hydrogenation and deuteration of $C_2H_2$ and $C_2H_4$ on cold grains: a clue to the formation mechanism of $C_2H_6$ with astronomical interest. *Astrophys. J.* **2017**, *837*, 155.

(65) Zhao, Y.; Truhlar, D. G. Hybrid meta density functional theory methods for thermochemistry, thermochemical kinetics, and non-covalent interactions: the MPW1B95 and MPWB1K models and comparative assessments for hydrogen bonding and van der Waals interactions. *J. Phys. Chem. A* **2004**, *108*, 6908−6918.

(66) Weigend, F.; Häser, M.; Patzelt, H.; Ahlrichs, R. RI-MP2: optimized auxiliary basis sets and demonstration of efficiency. *Chem. Phys. Lett.* **1998**, *294*, 143−152.

(67) Valiev, M.; Bylaska, E. J.; Govind, N.; Kowalski, K.; Straatsma, T. P.; van Dam, H. J.; Wang, D.; Nieplocha, J.; Apra, E.; Windus, T. L.; de Jong, W. NWChem: a comprehensive and scalable open-source solution for large scale molecular simulations. *Comput. Phys. Commun.* **2010**, *181*, 1477−1489.

(68) Zhao, Y.; Truhlar, D. G. The M06 suite of density functionals for main group thermochemistry, thermochemical kinetics, noncovalent interactions, excited states, and transition elements: two new functionals and systematic testing of four M06-class functionals and 12 other functionals. *Theor. Chem. Acc.* **2008**, *120*, 215−241.

(69) Knowles, P. J.; Hampel, C.; Werner, H.-J. Coupled cluster theory for high spin, open shell reference wave functions. *J. Chem. Phys.* **1993**, *99*, 5219−5227.

(70) Knowles, P. J.; Hampel, C.; Werner, H.-J. Erratum: Coupled cluster theory for high spin, open shell reference wave functions [J. Chem. Phys. 99, 5219 (1993)]. *J. Chem. Phys.* **2000**, *112*, 3106−3107.

(71) Deegan, M. J.; Knowles, P. J. Perturbative corrections to account for triple excitations in closed and open shell coupled cluster theories. *Chem. Phys. Lett.* **1994**, *227*, 321−326.

(72) Adler, T. B.; Knizia, G.; Werner, H.-J. A simple and efficient CCSD (T)-F12 approximation. *J. Chem. Phys.* **2007**, *127*, 221106.

(73) Peterson, K. A.; Adler, T. B.; Werner, H.-J. Systematically convergent basis sets for explicitly correlated wavefunctions: The atoms H, He, B−Ne, and Al−Ar. *J. Chem. Phys.* **2008**, *128*, 084102.

(74) Knizia, G.; Adler, T. B.; Werner, H.-J. Simplified CCSD (T)-F12 methods: theory and benchmarks. *J. Chem. Phys.* **2009**, *130*, 054104.

(75) Werner, H.; Knowles, P.; Knizia, G.; Manby, F.; Schütz, M. Molpro: a general-purpose quantum chemistry program package *WIREs Comput. Mol. Sci.* **2012**, *2*, 242−253.

(76) Kastner, J.; Carr, J. M.; Keal, T. W.; Thiel, W.; Wander, A.; Sherwood, P. DL-FIND: an open-source geometry optimizer for atomistic simulations. *J. Phys. Chem. A* **2009**, *113*, 11856−11865.

(77) Sherwood, P.; et al. QUASI: A general purpose implementation of the QM/MM approach and its application to problems in catalysis. *J. Mol. Struct.: THEOCHEM* **2003**, *632*, 1−28.

(78) Metz, S.; Kästner, J.; Sokol, A. A.; Keal, T. W.; Sherwood, P. C hem S hella modular software package for QM/MM simulations. *Wiley Interdiscip. Rev.: Comput. Mol. Sci.* **2014**, *4*, 101−110.

(79) Meisner, J.; Markmeyer, M. N.; Bohner, M. U.; Kaestner, J. Comparison of classical reaction paths and tunneling paths studied with the semiclassical instanton theory. *Phys. Chem. Chem. Phys.* **2017**, *19*, 23085−23094.

(80) Hratchian, H. P.; Schlegel, H. B. Accurate reaction paths using a Hessian based predictor−corrector integrator. *J. Chem. Phys.* **2004**, *120*, 9918−9924.

(81) Abplanalp, M. J.; Góbi, S.; Kaiser, R. I. On the formation and the isomer specific detection of methylacetylene ($CH_3CCH$), propene ($CH_3CHCH_2$), cyclopropane ($cC_3H_6$), vinylacetylene ($CH_2CHCCH$), and 1, 3-butadiene ($CH_2CHCHCH_2$) from interstellar methane ice analogues. *Phys. Chem. Chem. Phys.* **2019**, *21*, 5378−5393.

(82) Comeford, J.; Gould, J. H. Infrared spectra of solid hydrocarbons at very low temperatures. *J. Mol. Spectrosc.* **1961**, *5*, 474−481.

(83) Ghosh, J.; Hariharan, A. K.; Bhuin, R. G.; Methikkalam, R. R. J.; Pradeep, T. Propane and propane−water interactions: a study at cryogenic temperatures. *Phys. Chem. Chem. Phys.* **2018**, *20*, 1838−1847.

(84) Shaw, M. F.; Osborn, D. L.; Jordan, M. J.; Kable, S. H. Infrared spectra of gas-phase 1-and 2-propenol isomers. *J. Phys. Chem. A* **2017**, *121*, 3679−3688.

(85) Smith, R. S.; May, R. A.; Kay, B. D. Desorption kinetics of Ar, Kr, Xe, $N_2$, $O_2$, CO, methane, ethane, and propane from graphene and amorphous solid water surfaces. *J. Phys. Chem. B* **2016**, *120*, 1979−1987.

(86) Nykanen, L.; Honkala, K. Density functional theory study on propane and propene adsorption on Pt (111) and PtSn alloy surfaces. *J. Phys. Chem. C* **2011**, *115*, 9578−9586.

(87) Ioppolo, S.; Öberg, K.; Linnartz, H. In *Laboratory astrochemistry: from molecules through nanoparticles to grains*; Schlemmer, S., Giesen, T., Mutschke, H., Eds.; John Wiley & Sons: Weinheim, Germany, 2014; Chapter 5.4, pp 289−309.

(88) Collings, M. P.; Anderson, M. A.; Chen, R.; Dever, J. W.; Viti, S.; Williams, D. A.; McCoustra, M. R. A laboratory survey of the thermal desorption of astrophysically relevant molecules. *Mon. Not. R. Astron. Soc.* **2004**, *354*, 1133−1140.

(89) Burdett, J. L.; Rogers, M. T. Ketoenol tautomerism in β-dicarbonyls studied by nuclear magnetic resonance spectroscopy. III. studies of proton chemical shifts and equilibrium constants at different temperatures1. *J. Phys. Chem.* **1966**, *70*, 939−941.

(90) Isokoski, K.; Poteet, C.; Linnartz, H. Highly resolved infrared spectra of pure $CO_2$ ice (15−75 K). *Astron. Astrophys.* **2013**, *555*, A85.

(91) Schaff, J. E.; Roberts, J. T. The adsorption of acetone on thin films of amorphous and crystalline ice. *Langmuir* **1998**, *14*, 1478−1486.

(92) Dewick, P. M. *Essentials of organic chemistry: For students of pharmacy, medicinal chemistry and biological chemistry*; John Wiley & Sons: New York, 2006.

(93) Maity, S.; Kaiser, R. I.; Jones, B. M. Formation of complex organic molecules in methanol and methanol−carbon monoxide ices exposed to ionizing radiation−a combined FTIR and reflectron time-of-flight mass spectrometry study. *Phys. Chem. Chem. Phys.* **2015**, *17*, 3081−3114.

(94) Zaverkin, V.; Lamberts, T.; Markmeyer, M.; Kästner, J. Tunnelling dominates the reactions of hydrogen atoms with unsaturated alcohols and aldehydes in the dense medium. *Astron. Astrophys.* **2018**, *617*, A25.

(95) Arasa, C.; Andersson, S.; Cuppen, H.; van Dishoeck, E.; Kroes, G.-J. Molecular dynamics simulations of the ice temperature dependence of water ice photodesorption. *J. Chem. Phys.* **2010**, *132*, 184510.

(96) Fredon, A.; Lamberts, T.; Cuppen, H. Energy dissipation and nonthermal diffusion on interstellar ice grains. *Astrophys. J.* **2017**, *849*, 125.

(97) Vöhringer-Martinez, E.; Hansmann, B.; Hernandez, H.; Francisco, J.; Troe, J.; Abel, B. Water catalysis of a radical-molecule gas-phase reaction. *Science* **2007**, *315*, 497−501.

(98) Rimola, A.; Skouteris, D.; Balucani, N.; Ceccarelli, C.; Enrique-Romero, J.; Taquet, V.; Ugliengo, P. Can formamide be formed on interstellar ice? An atomistic perspective. *ACS Earth Space Chem.* **2018**, *2*, 720−734.

(99) Xu, S.; Lin, M. Theoretical study on the kinetics for OH reactions with CH3OH and C2H5OH. *Proc. Combust. Inst.* **2007**, *31*, 159−166.

(100) Gonzalez, J.; Caballero, M.; Aguilar-Mogas, A.; Torrent-Sucarrat, M.; Crehuet, R.; Solé, A.; Giménez, X.; Olivella, S.; Bofill, J. M.; Anglada, J. M. The reaction between HO and (H 2 O) n (n= 1, 3) clusters: reaction mechanisms and tunneling effects. *Theor. Chem. Acc.* **2011**, *128*, 579−592.

(101) Hargreaves, W. R.; Deamer, D. W. Liposomes from ionic, single-chain amphiphiles. *Biochemistry* **1978**, *17*, 3759−3768.

(102) Alamgir, A. *Therapeutic use of medicinal plants and their extracts: Phytochemistry and bioactive compounds*, Vol. 2; Progress in Drug Research, Vol. 74; Springer: Cham, Switzerland, 2018.

(103) Gerakines, P.; Moore, M. H.; Hudson, R. L. Carbonic acid production in $H_2O:CO_2$ ices. UV photolysis vs. proton bombardment. *Astron. Astrophys.* **2000**, *357*, 793−800.

(104) Cottin, H.; Moore, M. H.; Bénilan, Y. Photodestruction of relevant interstellar molecules in ice mixtures. *Astrophys. J.* **2003**, *590*, 874−881.